\documentclass[prl,twocolumn,floatfix,showpacs,superscriptaddress]{revtex4}
\usepackage{amsfonts}
\usepackage{amsmath}
\usepackage{amssymb}
\usepackage{graphicx}
\usepackage{natbib}
\usepackage[colorlinks=true]{hyperref}

\begin{document}

\title{Strong coupling superconductivity, pseudogap and Mott transition}
\author{G. Sordi}
\affiliation{Theory Group, Institut Laue Langevin, 6 rue Jules Horowitz, 38042 Grenoble Cedex, France}
\author{P. S\'emon}
\affiliation{D\'epartement de physique and Regroupement qu\'eb\'equois sur les mat\'eriaux de pointe, Universit\'e de Sherbrooke, Sherbrooke, Qu\'ebec, Canada J1K 2R1}
\author{K. Haule}
\affiliation{Department of Physics \& Astronomy, Rutgers University, Piscataway, NJ 08854-8019, USA}
\author{A.-M. S. Tremblay}
\affiliation{D\'epartement de physique and Regroupement qu\'eb\'equois sur les mat\'eriaux de pointe, Universit\'e de Sherbrooke, Sherbrooke, Qu\'ebec, Canada J1K 2R1}
\affiliation{Canadian Institute for Advanced Research, Toronto, Ontario, Canada, M5G 1Z8}
\pacs{71.27.+a, 71.10.Fd, 71.10.Hf, 71.30.+h}

\date{\today}

\begin{abstract}
An intricate interplay between superconductivity, pseudogap and Mott transition, either bandwidth driven or doping driven, occurs in materials. Layered organic conductors and cuprates offer two prime examples.
We provide a unified perspective of this interplay in the two-dimensional Hubbard model within cellular dynamical mean-field theory on a $2\times 2$ plaquette and using the continuous-time quantum Monte Carlo method as impurity solver.
Both at half filling and at finite doping, the metallic normal state close to the Mott insulator is unstable to d-wave superconductivity. Superconductivity can destroy the first-order transition that separates the pseudogap phase from the overdoped metal, yet that normal state transition leaves its marks on the dynamic properties of the superconducting phase. For example, as a function of doping one finds a rapid change in the particle-hole asymmetry of the superconducting density of states.
In the doped Mott insulator, the dynamical mean-field superconducting transition temperature $T_c^d$ does not scale with the order parameter when there is a normal-state pseudogap. $T_c^d$ corresponds to the local pair formation temperature observed in tunneling experiments and is distinct from the pseudogap temperature.
\end{abstract}
\maketitle

The proximity between a Mott insulator and a superconductor is one of the most intriguing puzzles in condensed matter physics~\cite{Anderson:1987}. Indeed, in a Mott insulator, strong Coulomb repulsion between electrons is at the origin of the phenomenon, while superconductivity is usually associated with effective attraction. In half-filled band layered organic conductor, pressure induces a first-order transition between a d-wave superconductor and a Mott insulator. This is a bandwidth-induced transition.
The maximum superconducting transition temperature $T_c$ is at the first-order phase boundary~\cite{Lefebvre:2000}.
On the contrary, in high-temperature superconductors, while superconductivity emerges upon doping a Mott insulator, $T_c$ has a dome shape and disappears before the doping driven Mott transition~\cite{ift}. In addition, the normal state near the Mott insulator exhibits a pseudogap~\cite{normanADV}.

Weak coupling approaches to the simplest model that includes screened Coulomb interaction and band structure effects, the Hubbard model, show that d-wave superconductivity can arise as a secondary effect from exchange of antiferromagnetic fluctuations~\cite{Beal-Monod:1986,Scalapino:1986,HalbothPom:2000,Kyung:2003,Bourbonnais:2009,ScalapinoThread:2010}. At strong coupling, renormalized mean-field theory~\cite{Miyake:1986,AndersonVanilla:2004,Yang:2006}, slave particle~\cite{LeeRMP:2006,Imada:2011} and variational approaches~\cite{Giamarchi:1991,Paramekanti:2004} also suggest the presence of d-wave superconductivity. However, to study both the Mott transition and d-wave superconductivity, one must resort to cluster versions of dynamical mean-field theory~\cite{Hettler:1998,gabiCDMFT,maier,kotliarRMP}. Up to now, results have been obtained mostly at zero temperature~\cite{davidAF,kyung:2006,AichhornAFSC:2006,massimoAF,markus,kancharla,civelli1,Balzer:2010,Hanke:2010,Weber:2011}. There are also a few results on the transition temperature~\cite{maierAF,maierSystem,hauleDOPING,sentefSC} but there is no systematic study of the interplay of superconductivity and pseudogap with both bandwidth-driven and doping-driven Mott transitions at finite temperature. This is the problem that we solve in this Letter by studying the two-dimensional Hubbard model with cellular dynamical mean-field theory on a plaquette~\cite{maier,kotliarRMP} using state of the art continuous-time Quantum Monte Carlo method as impurity solver~\cite{millisRMP,Werner:2006,WernerCTQMC,hauleCTQMC}. 
Notice that quite generally~\cite{mermin-wagner} there is no continuous symmetry breaking in two dimensions at finite temperature. This is true for d-wave superconductivity as well~\cite{su-suzuki}. However, it is still physically meaningful to study the superconducting phase at the dynamical mean-field level since the corresponding transition temperature $T_c^d$ indicates where the superconducting fluctuations begin to develop. Three-dimensional effects eventually allow true long-range order at lower temperature. 
Competition with other long-range ordered phases~\cite{Fradkin:2010}, which are influenced by many factors including frustration, will be considered in future work.

After we present the model and method, we discuss in turn the bandwidth-driven and the doping-driven cases before we provide a unified view and discussion of the results.

%
{\it Model and method.}--
We consider the two-dimensional Hubbard model on a square lattice,
\begin{equation}
  H=-\sum_{ij\sigma}t_{ij}c_{i\sigma}^\dagger c_{j\sigma}
  +U\sum_{i}\left(n_{i\uparrow }-1/2\right)\left(n_{i\downarrow }-1/2\right)
  -\mu\sum_{i\sigma} n_{i\sigma}
\label{eq:HM}
\end{equation}
where $c^{+}_{i\sigma}$ and $c_{i\sigma}$ create and annihilate an electron of spin $\sigma$ on site $i$, $n_{i\sigma}=c^{+}_{i\sigma}c_{i\sigma}$, $t$ is the nearest neighbor hopping amplitude, $\mu$ is the chemical potential and $U$ is the screened Coulomb repulsion.
We solve this model using cellular dynamical mean-field theory (CDMFT)~\cite{kotliarRMP,maier}.
This approach takes a cluster of lattice sites, here a $2\times 2$ plaquette, out of the lattice, and embeds it in a self-consistent bath of noninteracting electrons.
The action of the plaquette coupled to the bath reads
\begin{equation}
  S = S_{c} +\int_{0}^{\beta} d\tau \int_{0}^{\beta} d\tau' {\bf \psi^{\dag}}(\tau) \hat{\Delta}(\tau,\tau') {\bf \psi}(\tau') ,
\label{eq:action}
\end{equation}
where $S_{c}$ is the action of the cluster and $\hat{\Delta}$ the hybridization matrix. From now on, the symbol $\hat{\ }$ indicates a matrix in cluster indices.
The hybridization $\hat{\Delta}$ is determined by the self-consistency condition
\begin{equation}
\hat{\Delta}(i\omega_{n}) =  i\omega_{n} +\mu -\hat{t}_{c}-\hat{\Sigma}_{c}(i\omega_{n}) -\hat{G}(i\omega_n)^{-1}
\label{eq:SCC}
\end{equation}
which states that the infinite lattice and plaquette have the same self-energy and the same Green's function on the plaquette.
Here $\hat{\Sigma}_{c}$ is the cluster self-energy, $\hat{t}_{c}$ the cluster hopping, and $\hat{G}(i\omega_n)=\sum_{\tilde{k}} \frac{1}{i\omega_{n} +\mu -\hat{t}(\tilde{k}) -\hat{\Sigma}_{c}(i\omega_{n})}$, where $\tilde{k}$ is the superlattice momentum. 
We solve the impurity (plaquette+bath) problem of Eq.~(\ref{eq:action}) using the continuous-time quantum Monte Carlo method~\cite{millisRMP,hauleCTQMC}, which sums all diagrams obtained by the expansion of the action of Eq.~(\ref{eq:action}) with respect to the hybridization $\hat{\Delta}$.
For the superconducting state in the cluster momentum basis, the cluster Nambu Green's function reads
\begin{equation}
G_{K}(\tau) = \left( \begin{array}{cc}
					G_{K\uparrow}(\tau) & F_{K}(\tau) \\
					F_{K}^{+}(\tau) & -G_{-K\downarrow}(-\tau)
                	\end{array} \right)
\label{eq:Green}
\end{equation}
where $F$ is the anomalous Green's function.
For d-wave superconductivity, $F_{(\pi,0)}=-F_{(0,\pi)}$ is the only nonzero component.
To determine the parameter space where the superconducting phase is allowed by the CDMFT equations, we monitor the superconducting order parameter $\Phi=\langle F_{(\pi,0)}(\tau=0^{+}) \rangle$.

\begin{figure}[!ht]
\centering{
\includegraphics[width=0.90\linewidth,clip=]{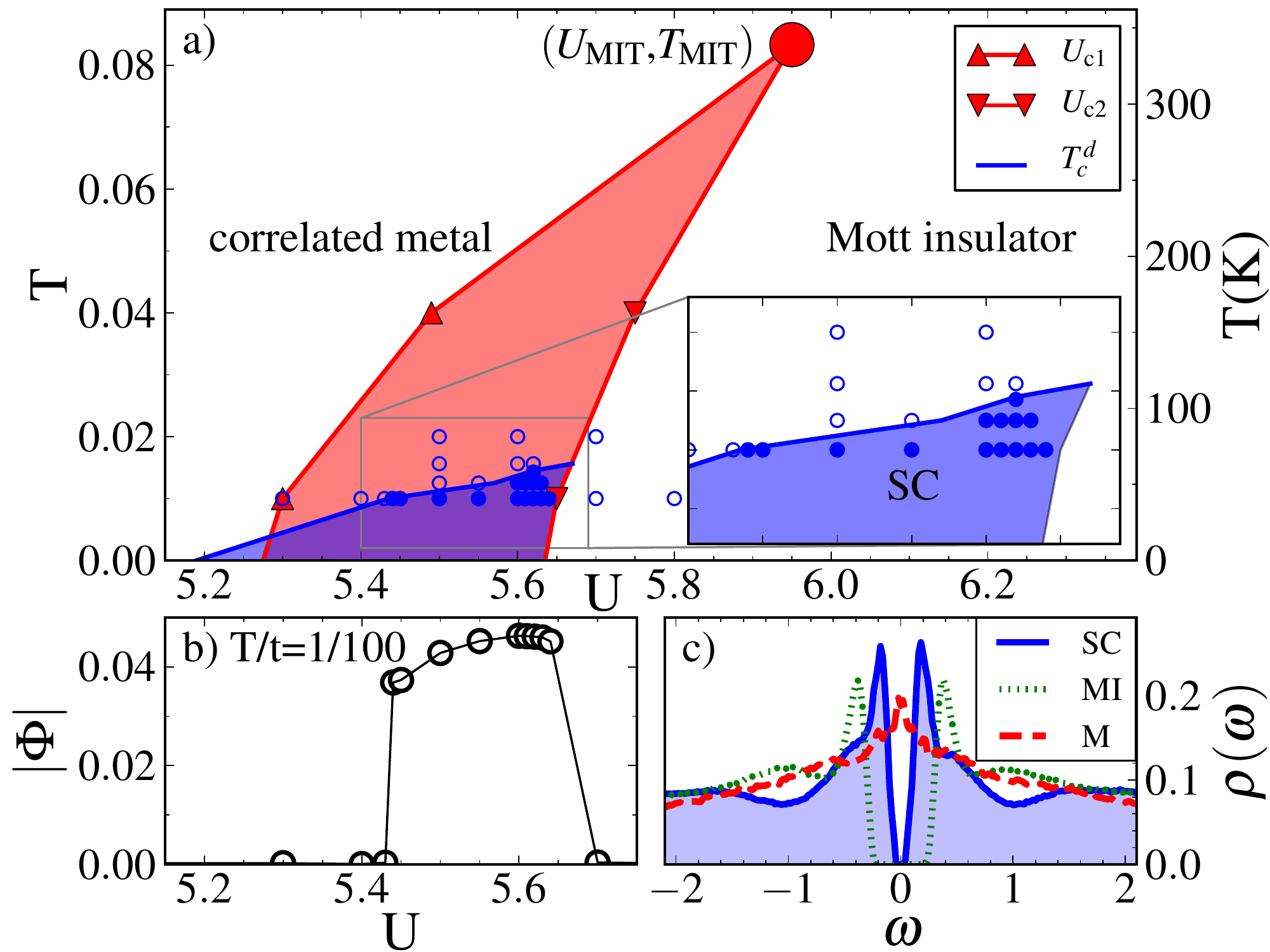}}
\caption{(a) Temperature $T$ versus interaction strength $U$ phase diagram of the half-filled two-dimensional Hubbard model obtained by CDMFT. Three phases can be distinguished: correlated metal, Mott insulator, and superconductor. In the normal state, there is a first-order transition at finite temperature between a correlated metal and a Mott insulator, bounded by the spinodals $U_{\rm c2}(T)$ and $U_{\rm c1}(T)$, defined as the loci where the double occupation shows a jump. The superconducting phase (blue or dark gray region) is defined by the loci where $|\Phi|\neq 0$ (filled blue circles) and is delimited by the superconducting transition temperature $T_{c}^{d}$. Extrapolations to $T=0$ are a guide to the eye. On the right vertical axis we convert to physical units by using t = 0.35eV. Inset: zoom on the superconducting phase. (b) d-wave superconducting order parameter $\Phi$ as a function of $U$ at half filling and for $T/t=1/100$. (c ) Density of states $\rho(\omega)$ for $U=5.6t$ and $T/t=1/100$ for the normal-state Mott insulator, the normal-state metal and the superconductor (dotted, dashed, and solid lines, respectively).
}
\label{fig2}
\end{figure}

{\it Superconductivity and interaction-driven Mott transition.}--First, consider the normal state of the half-filled two dimensional Hubbard model.
Previous work revealed a first-order transition at moderate interaction between a correlated metal and a Mott insulator~\cite{phk,balzer,werner8}.
As shown in Fig.~\ref{fig2}a, in the $(U,T)$ plane there is a hysteresis region (in red or light gray) where two mean-field solutions can be obtained.
This region is bounded by the spinodals $U_{\rm c1}(T)$ and $U_{\rm c2}(T)$ (red lines with triangles) where the double occupation shows sudden jumps.
The first-order metal-insulator transition lies within this region and starts at the critical Mott endpoint $(U_{\rm MIT},T_{\rm MIT})\approx(5.95t,1/12t)$.

Next we allow for d-wave symmetry breaking in the CDMFT equations and perform scans as a function of $U$ for different temperatures.
As input seed of the CDMFT iterative procedure we use the normal state converged solution, and we add a small perturbation in the anomalous component of the hybridization matrix.
We obtain a converged superconducting solution, characterized by a nonzero $\Phi$, close to the Mott transition.
No superconducting solution is found if we use the metastable insulating solution as seed.
Figure~\ref{fig2}b shows the order parameter $\Phi$ for the low temperature $T/t=1/100$. 
Within our numerical precision, as a function of $U$, the order parameter exhibits two jumps: one at $U(T/t=1/100)\approx 5.45$ where there is a transition from the metal to the superconductor, and one at $U_{\rm c2}(T/t=1/100)\approx 5.65$ where the transition is between the superconductor and the insulator.

By performing the above procedure for different temperatures, we obtain the superconducting region in the $(U,T)$ plane [blue or dark gray region in Fig.~\ref{fig2}a], defined as the region where $\Phi \neq 0$.
With decreasing temperature, the superconducting phase emerges from the normal state metal close to the Mott transition, i.e. for $U<U_{\rm c2}$, and rapidly disappears below $U_{\rm c1}$.
The largest superconducting transition temperature $T_{c}^{d}(U)$ occurs, along with the largest order parameter, around the first-order boundary with the insulator, as in the organics~\cite{Lefebvre:2000}.

Physically, the CDMFT superconducting transition temperature $T_{c}^{d}$ is the temperature below which Cooper pairs form within the cluster. Previous work~\cite{maierSystem} suggests that $T_{c}^{d}$ converges to a finite value with cluster sizes up to 26 sites.
Long-wavelength thermal and quantum fluctuations in the magnitude~\cite{Ussishkin:2002} and phase of the order parameter~\cite{ekPRL,ekNat,Podolsky:2007,Tesanovic:2008} will lead to an {\it actual} superconducting transition temperature $T_{c}$ smaller than $T_{c}^{d}$. Long-wavelength antiferromagnetic fluctuations on the other hand can increase $T_{c}^{d}$, as seen in weak-coupling calculations~\cite{Beal-Monod:1986,Scalapino:1986,HalbothPom:2000,Kyung:2003,Bourbonnais:2009}. Competing long-range order would reduce or eliminate $T_{c}^{d}$~\cite{Fradkin:2010}.
Nevertheless, $T_{c}^{d}$ informs us on the regime of temperature where strong coupling and short-range nonlocal correlations lead to pairing.
These effects lead to a strong d-wave pairing gap in the density of states of Fig.~\ref{fig2}c.

\begin{figure}[t]
\centering{
\includegraphics[width=0.90\linewidth,clip=]{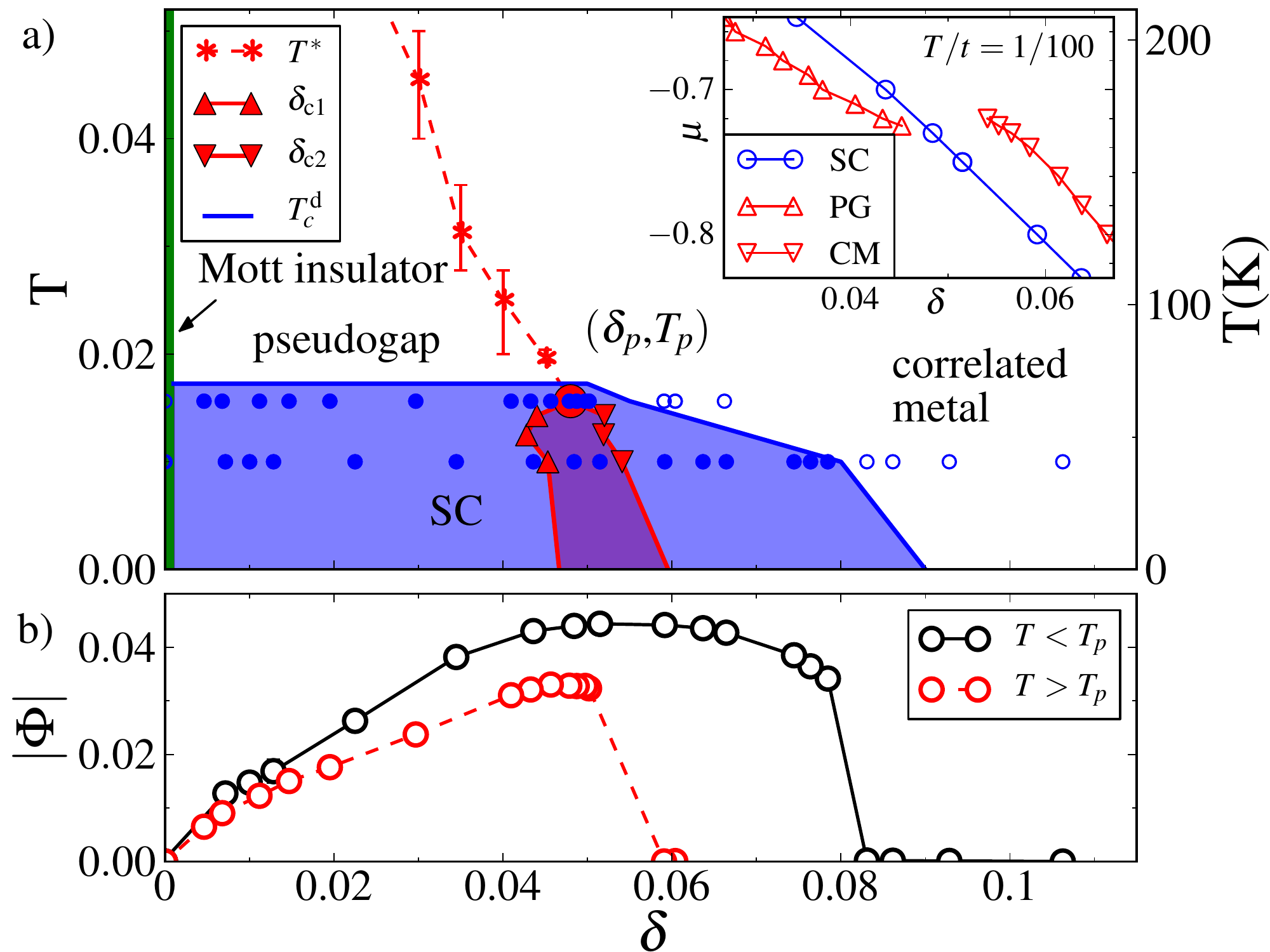}}\caption{(a) Temperature $T$ versus doping $\delta$ phase diagram  at $U=6.2t>U_{\rm MIT}$, obtained by CDMFT. Four phases can be recognized: in the normal state, there is a first-order transition at finite temperature between a pseudogap and a correlated metal, bounded by the spinodals $\delta_{\rm c1}(T)$ and $\delta_{\rm c2}(T)$ (up and down triangles respectively). A crossover takes place above the critical endpoint $(\delta_p,T_p)$ and defines the pseudogap temperature $T^{*}$~\cite{ssht}, determined by max $d\rho(\omega=0)/dT$. The third phase is the Mott insulator at $\delta=0$ (green solid line). The fourth phase, the superconducting one, is delimited by $T_{c}^{d}(\delta)$, i.e. the temperature below which $|\Phi|\neq 0$. Extrapolations to $T=0$ are a guide to the eye. Inset: chemical potential $\mu$ versus doping $\delta=1-n$ at $T=1/100$ for the normal state (triangles) and the superconducting state (circles). The jump in the dopings identify the spinodal points between the two normal-state metals, i.e. the pseudogap (PG) and the correlated metal (CM). The transition is removed by the superconducting state: $\mu(\delta)$ does not show any sign of hysteresis. (b) d-wave superconducting order parameter $\Phi$ as a function of doping for temperatures $T=1/64>T_{\rm p}$ and $1/100<T_{\rm p}$. On the right vertical axis we convert to physical
units by using t = 0.35eV.
}
\label{fig3}
\end{figure}
{\it Superconductivity and doping-driven Mott transition.}--We turn to the doped Mott insulator.
Previously, we explored the normal-state phase diagram~\cite{sht,sht2} and demonstrated that the first-order transition at half filling naturally extends at finite doping, and that it can take place between two metallic states: a correlated metal at large doping and a pseudogap~\cite{ssht}.
Figure~\ref{fig3}a shows the $(\delta,T)$ plane at $U=6.2t>U_{\rm MIT}$.
The spinodals $\delta_{\rm c1}(T)$ and $\delta_{\rm c2}(T)$, determined by the jumps in the doping $\delta$ (see inset), envelop the transition and terminate at the critical point $(\delta_p,T_p)$, which is the extension of the Mott critical point away from half filling. The value of $(\delta_p,T_p)$ moves to larger dopings and smaller temperatures as $U$ increases. At $U=6.2t$, $T_p$ is sufficiently large to be accessible by simulations.
Associated with the critical point $(\delta_p,T_p)$ there is a Widom line~\cite{water1}, and the pseudogap temperature $T^*(\delta)$ occurs along this line~\cite{ssht}.

Next, we study the superconducting phase as a function of doping.
The superconducting order parameter is shown in Fig.~\ref{fig3}b for different low temperatures.
In the Mott insulator at zero doping, $\Phi=0$ and thus there is no superconductivity.
Upon hole doping, $\Phi$ increases, reaches a maximum for a doping near the normal-state first-order transition between the pseudogap and correlated metal, and, with further doping, decreases.

By monitoring $\Phi(\delta)$ for different temperatures, we can construct the superconducting region in the $(\delta,T)$ plane (blue/dark grey region in Fig.~\ref{fig3}a).
The transition temperature $T_{c}^{d}$ is higher than the  critical temperature $T_{\rm p}$, and superconductivity eliminates the first-order transition of the underlying normal state. Indeed, the $\delta(\mu)$ curve in the inset of Fig.~\ref{fig3}a) is continuous.
$T_{c}^{d}$ is zero at $\delta=0$, but it is finite for $\delta\rightarrow 0^+$ and does not show large variation when there is a pseudogap in the underlying normal state.
In particular, $T_{c}^{d}(\delta)$ does not appreciably increase as we approach half filling while the pseudogap temperature $T^{*}$ does, showing that the two phenomena are distinct, as also found in high-field transport measurements ~\cite{Alloul:2010,Alloul:2011}.
With further doping, when the superconductivity evolves from a correlated metal, $T_{c}^{d}$ decreases and vanishes at large doping.
Therefore, our results imply that Mott physics causes $\Phi$ to drop at small doping, but does {\it not} produce a fall in $T_{c}^{d}$.
$T_{c}^{d}$ corresponds to Cooper pair formation within the plaquette. We associate $T_c^d$ to the temperature at which a superconducting gap appears in tunneling experiments~\cite{Gomes:2007,Gomes:2008} without long-range phase coherence. Experimentally, in the doping range where there is a normal-state pseudogap, that temperature scale is smaller than $T^*$ and larger than the actual $T_{c}$.
The small value of $\Phi$ suggests that the actual $T_{c}$ of the system will vanish at small doping due to competing order~\cite{Fradkin:2010} or to disorder~\cite{AlbenqueAlloul:2008,AlloulRMP:2009} or to long wavelength (classical and quantum) fluctuations of the magnitude~\cite{Ussishkin:2002} or of the phase~\cite{ekNat,ekPRL,Podolsky:2007,Tesanovic:2008} of the order parameter.

\begin{figure}[t]
\centering{
\includegraphics[width=0.90\linewidth,clip=]{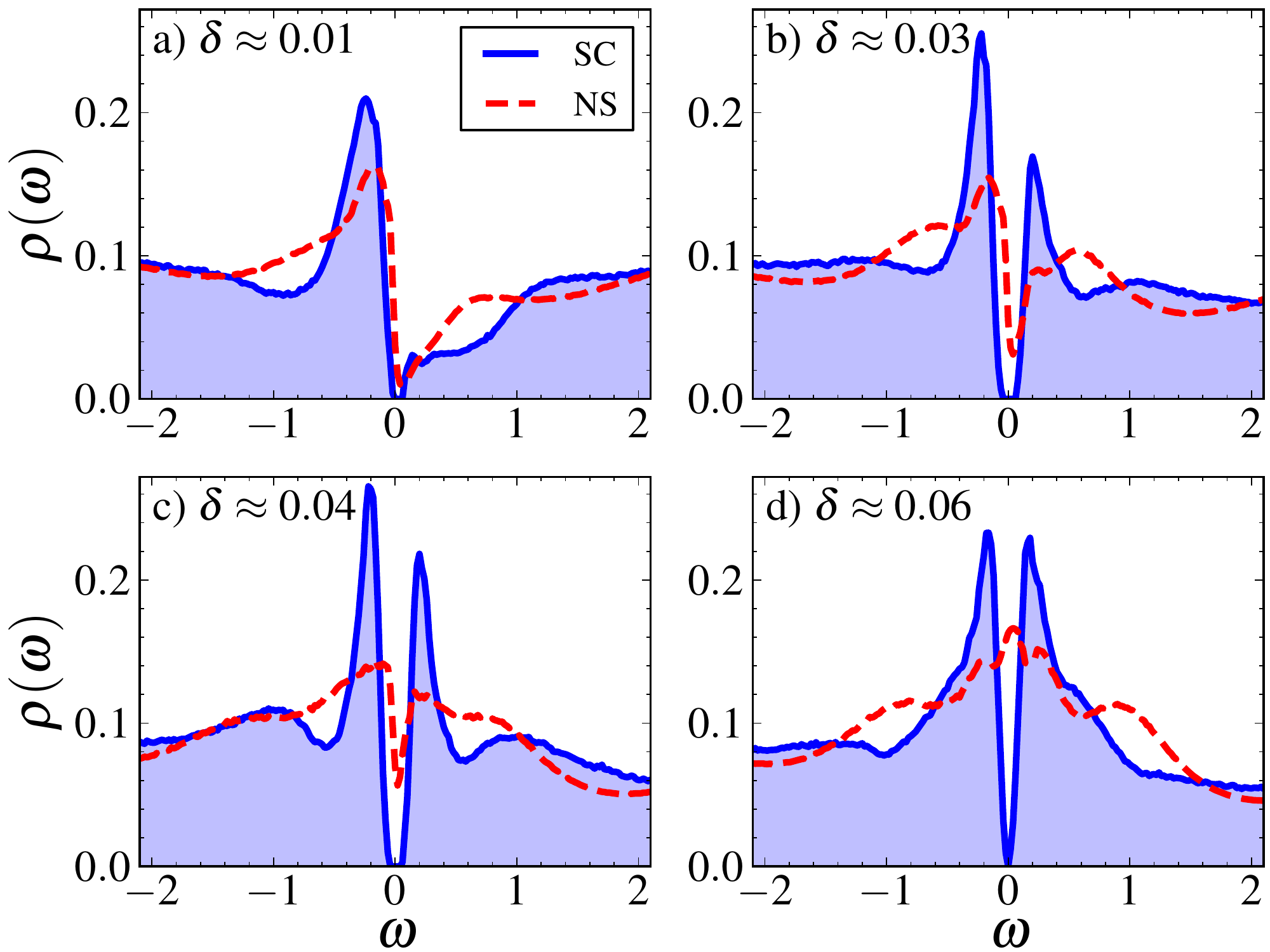}}\caption{Low frequency part of the local density of states $\rho(\omega)$ at $U=6.2t$, $T/t=1/100$ for the normal-state and the superconducting state (red dashed and blue solid lines). For $\delta\approx 0.01, 0.03, 0.04$ [panels a, b, c respectively] the superconducting state emerges from the underlying normal-state pseudogap metal. It inherits a strong particle-hole asymmetry. For $\delta\approx 0.06$ the superconducting state emerges from a correlated normal-state metal, and the density of states, near the transition, approximately recovers particle-hole symmetry at low frequency.
}
\label{fig4}
\end{figure}
Even though superconductivity eliminates the first-order transition in the underlying normal state, signatures of that transition remain in the dynamics of the superconducting state.
This is shown by the evolution of the density of states with doping in Fig.~\ref{fig4} where the solid line is for the superconducting state and the dashed line for the normal state.
At low doping, superconductivity originates from the pseudogap and the superconducting density of states inherits its large particle-hole asymmetry~\cite{hauleDOPING}, as found in experiments~\cite{FischerRMP:2007};
On the other side of the transition, at large doping, superconductivity emerges from the normal-state correlated metal, and the superconducting density of states at low frequency close to the normal-state transition is particle-hole symmetric.
Our contribution is to link the features of the superconducting density of states to the underlying normal-state first-order transition.

%
\begin{figure}[!t]
\centering{
\includegraphics[width=0.90\linewidth,clip=]{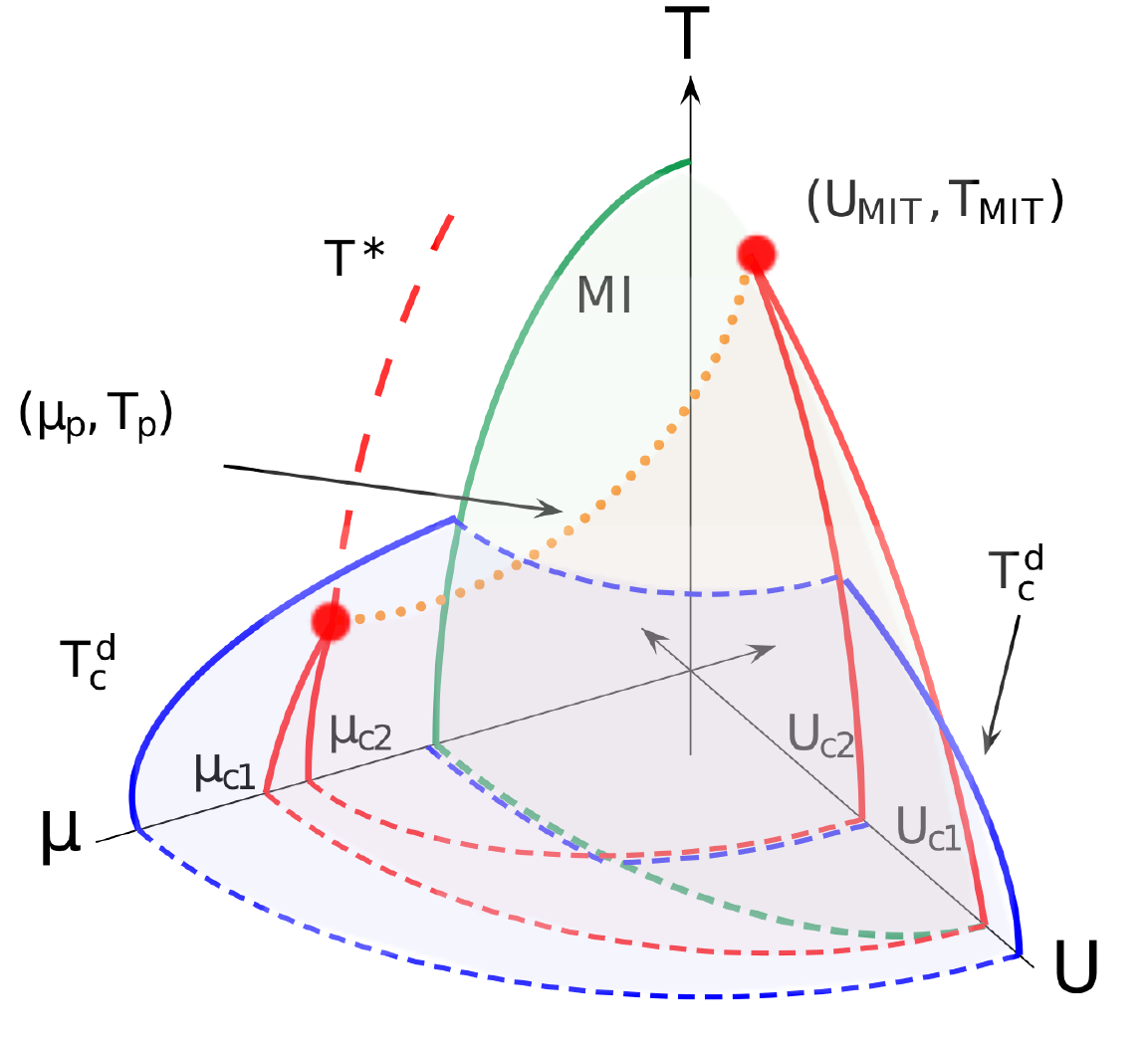}}
\caption{Schematic temperature - chemical potential - interaction strength phase diagram based on CDMFT solution of the 2D Hubbard model. Cut at particle-hole symmetry ($\mu=0$) and at constant $U>U_{\rm MIT}$ are shown. Since we set $t'=0$, the phase diagram is symmetric with respect to $\mu=0$ plane. The first-order transition between a metal and a Mott insulator in the $\mu=0$ plane is connected with the first-order transition between the pseudogap and a correlated metal in the $U>U_{\rm MIT}$ plane~\cite{sht,sht2}. $T_{p}$ begins at $T_{\rm MIT}$. The superconducting temperature $T_{c}^{d}$, delimiting the region where $\Phi$ is non zero, is also shown. In the phase diagram, the superconducting phase emerges from the normal state metal close to the Mott insulator.
}
\label{fig1}
\end{figure}

{\it Superconductivity from Mott physics.}--The above analysis shows that superconductivity arises by approaching the Mott insulator as a function of both the interaction strength and the doping.
The two routes to create superconductivity are related, as sketched by the $(U,\mu,T)$ phase diagram in Fig.~\ref{fig1}. The critical endpoint $(\mu_{p},T_{p})$, hidden by the superconducting phase in the $(\mu,T)$ plane of that figure, is connected to the familiar Mott endpoint $(U_{\rm MIT}, T_{\rm MIT})$ at half filling (see dotted line in Fig.~\ref{fig1}). The latter appears above the superconducting phase.
Recent works at half filling~\cite{sentefSC} did not find a direct transition between the superconductor and the Mott insulator.

Previous CDMFT works~\cite{tremblayR,kancharla,massimoAF,hauleDOPING,civelli1,Balzer:2010,Weber:2011} at zero temperature reported a doping dependence of the order parameter $\Phi$ similar to the one found here, but the doping dependence of $T_{c}^{d}$ could only be surmised.
Our contribution is to show that $T_{c}^{d}$ does not scale with $\Phi(\delta)$ when a pseudogap is the underlying normal state.
$T_{c}^{d}$ remains finite as the Mott insulator is approached, implying that Mott physics does not suppress $T_{c}^{d}$ even though it suppresses the order parameter. In the region where there is a normal-state pseudogap, $T_{c}^{d}$ represents a local pair-formation~\cite{Gomes:2007,Gomes:2008} temperature scale that is distinct from both $T^*$ and the actual superconducting long-range phase coherence $T_c$.
In addition, we find that a classical, not quantum, critical point at finite temperature between a pseudogap and a correlated metal~\cite{sht,sht2,ssht}, continues to control the distinct pseudogap physics above $T_{c}^{d}$, even though the superconducting phase replaces the normal-state first-order transition at low temperature.
This finding has to be contrasted with the quantum critical point reported in previous work~\cite{jarrellNFL}.
Because those calculations were limited to high temperatures, they did not detect the normal-state first-order transition.

The phase diagram as a function of interaction strength, doping and temperature that we found shows that a transition directly to the superconducting state from a Mott insulator is possible at the dynamical mean-field level, whether the transition is bandwidth or doping driven. Since $T_{c}^{d}$ is finite at infinitesimal doping, the transition appears as first-order in both cases. Hence, the experimentally observed drop of $T_c$ at low doping must come from mechanisms not included here, such as long wavelength fluctuations~\cite{ekPRL,ekNat,Ussishkin:2002,Podolsky:2007,Tesanovic:2008}, competing order~\cite{Fradkin:2010} or disorder~\cite{AlbenqueAlloul:2008,AlloulRMP:2009}. Long-wavelength fluctuations should be important near the Mott transition because the order parameter decreases rapidly with decreasing doping, contrary to $T_{c}^{d}$. Yet, $T_{c}^{d}$ retains a role as a local pair formation temperature~\cite{Gomes:2007,Gomes:2008} and is distinct from the pseudogap temperature $T^*$. For sufficiently large $U$ the superconducting state destroys the underlying first-order transition between the pseudogap and the correlated metal, but signatures of this transition remain in the dynamical properties of the superconductor.

We are indebted to S. Allen for technical help. We thank E. Kats, T. Ziman, A. Cano and G. Kotliar for useful discussions.
This work was partially supported by FQRNT, by the Tier I Canada Research Chair Program (A.-M.S.T.), and by NSF DMR-0746395 (K.H.). A.-M.S.T is grateful to the Harvard Physics Department for support and P.S. for hospitality during the writing of this work. Partial support was also provided by the MIT-Harvard Center for Ultracold Atoms. Simulations were performed on computers provided by CFI, MELS, Calcul Qu\'ebec and Compute Canada. Portions of the hybridization expansion impurity solver developed by P.S. were based on the ALPS library~\cite{ALPS}.


\end{document}